# Ionization of protoplanetary disks by galactic cosmic rays, solar protons, and by supernova remnants

Ryuho Kataoka (1,2), and Tatsuhiko Sato (3)

(1) National Institute of Polar Research, 10-3 Midori-cho, Tachikawa, Tokyo, 190-8513 Japan.
(2) SOKENDAI, Japan. (Email: kataoka.ryuho@nipr.ac.jp)
(3) Japan Atomic Energy Agency, Shirakata 2-4, Tokai, Ibaraki 319-1195, Japan.

**Abstract**

Galactic cosmic rays and solar protons ionize the present terrestrial atmosphere, and the air showers are simulated by well-tested Monte-Carlo simulations, such as PHITS code. We use the latest version of PHITS to evaluate the possible ionization of protoplanetary disks by galactic cosmic rays (GCRs), solar protons, and by supernova remnants. The attenuation length of GCR ionization is updated as 118 g cm$^{-2}$, which is approximately 20% larger than the popular value. Hard and soft possible spectra of solar protons give comparable and 20% smaller attenuation lengths compared with those from standard GCR spectra, respectively, while the attenuation length is approximately 10% larger for supernova remnants. Further, all of the attenuation lengths become 10% larger in the compound gas of cosmic abundance, e.g. 128 g cm$^{-2}$ for GCRs, which can affect the minimum estimate of the size of dead zones in protoplanetary disks when the incident flux is unusually high.

## 1. Introduction

Galactic cosmic rays (GCRs) are the major ionization sources of the present terrestrial atmosphere. One of the best-tested GCR-induced air shower simulation codes is PHITS: Particle and Heavy Ion Transport code System (Sato, 2013a). The code has been employed in GCR-induced air shower simulation for developing the PARMA model, which comprises numerous analytical functions with parameters whose numerical values were fitted to reproduce the PHITS simulation results (Sato, 2015). PHITS has also been used in solar flare-related air shower simulation for developing WASAVIES: WArning System for AVIation exposure to Solar energetic particles (Kataoka et al., 2014a; 2015, Sato 2013b). Recently, Kataoka et al. (2014b) showed an extreme example of PHITS's applications to the expected radiation dose at the Earth during the catastrophic collision of the heliosphere with a nearby supernova remnant as simulated by Fields et al. (2008).

A star formation may also be affected by the extremely enhanced ionization of a molecular cloud due to a supernova remnant, via changing both the ambipolar diffusion time and the distribution of dead zones in the accretion disk (Fatuzzo et al., 2006), and can also be an important target of PHITS simulations. In a "dead zone" of protoplanetary disks (Gammie, 1996), presumably near the midplane where ionization level is the lowest, the turbulence resulting from magneto-rotational instability (Balbus and Hawley, 1991) cannot participate for the angular momentum transport, which has been regarded as the key to the evolution of protoplanetary disks themselves, dust grains, and therefore the origin of planets (Sano et al., 2000; Okuzumi et al., 2012). Turner and Drake (2009) suggested that a variety of dead zones can exist in protoplanetary disks, considering the ionizations from enhanced GCRs or from enhanced solar protons. In fact, T Tauri stars can be magnetically more active than the present Sun to truncate the accretion disk by the dipole-like magnetic field (Shu et al., 1994), and solar protons from the young Sun can have a significant impact on the ionizations of the protoplanetary disk, which can be comparable with those by GCRs.

The most popular model of GCR-induced ionization profile for protoplanetary disks was proposed 35 years ago by Umebayashi and Nakano (1981), hereafter UN81, and the attenuation length of GCR-induced ionization was estimated as 96 g cm$^{-2}$ for a pure hydrogen gas. The purpose of this study is to examine and update the attenuation length for a pure hydrogen gas and also for a compound gas of cosmic abundance using the latest version of PHITS code. We discuss the possible range of the variable ionization profiles due to other energetic particles with softer or harder spectra, such as solar protons from the young Sun and also energetic protons generated at supernova remnants, respectively. A new model of proton spectra of the young Sun is also proposed for this purpose, as a function of magnetic moment of the central star and mass loss rate.

## 2. Method

We consider cylindrical areas with homogeneous gas density of $10^{10}$ cm$^{-3}$ and $10^{14}$ cm$^{-3}$. Two different compositions are selected, i.e. pure hydrogen gases and compound gases of cosmic abundance with 75% hydrogen and 25% helium in the mass ratio. The height and radius of the cylindrical areas are fixed to 1500 g cm$^{-2}$. We assume that the GCRs or solar protons with different energy spectra (**Figure 1**) are perpendicularly incident to the cylindrical gas at the center of the flat top surface. The deposition energy is then scored as a function of the depth from the front surface.

The latest version of PHITS, v.2.82 with the recommended nuclear reaction models and data libraries such as JAM (Nara et al., 2000), INCL4.6 (Boudard et al., 2013) and EGS5 (Hirayama et al., 2005), were employed in the simulation. The decay of neutrons is indispensable to be considered in the simulation because of the extremely large scale of the simulation phase space, up to $10^{12}$ km for the lowest density case. However, PHITS 2.82 cannot consider the decay of neutrons because the mean life of neutron is so long in comparison to the time scale of conventional particle transport simulation. In this study we therefore implement the new function to consider the neutron decay with the mean life time of 887 sec. This function will be available in the forthcoming version of PHITS. Note that the decay of other particles such as pions and muons can be taken into account in the default setting of PHITS 2.82.

Energy spectra of GCRs at the present Earth position are given as a reference in this study by the model developed by Matthiä et al. (2013), as a function of solar modulation index $W$. We use the minimum ($W = 0$) and maximum ($W = 200$) modulation indices to see the range of the weakest and strongest current solar activities, respectively. The same GCR spectrum as used in UN81 is also used as another reference. Note that Cleeves et al. (2013) suggested T-Tauliosphere modulation model, which consider stronger GCR modulation than the maximum modulation assumed in this paper. The absolute values of the all energy spectra adopted in this study are normalized to give the same flux of UN81 at 10 GeV, as shown in **Figure 1**.

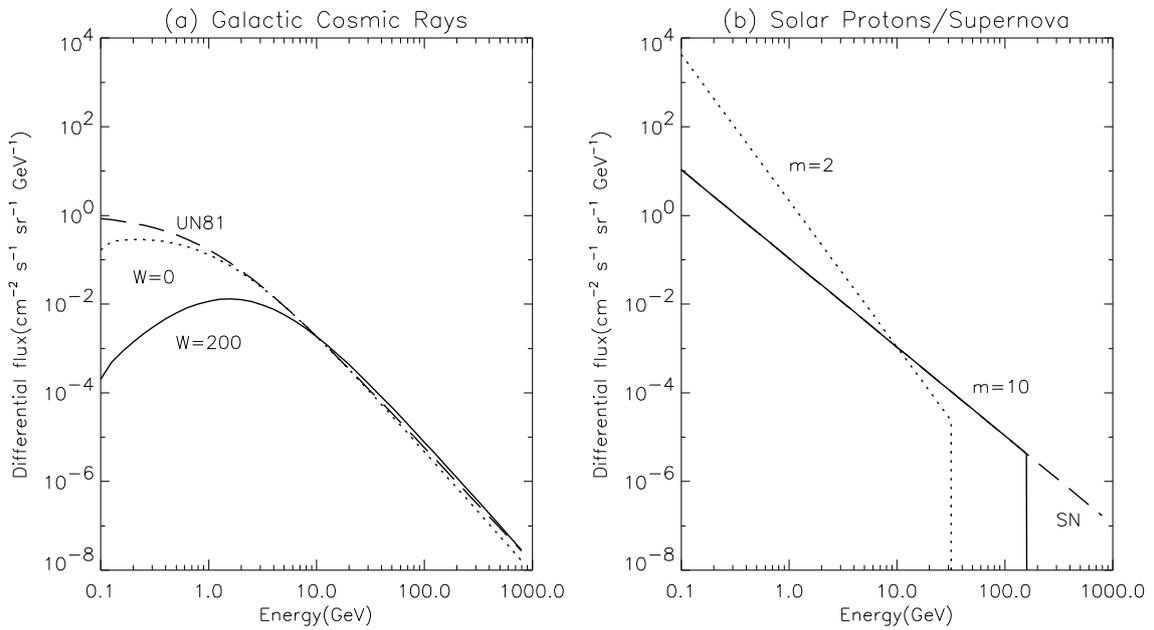

**Figure 1:** Assumed energy spectra for (a) galactic cosmic rays of weak ($W = 0$) and strong ($W = 200$) solar activities and of UN81 model, and for (b) solar protons of weak shock ($m = 2$), strong shock ($m = 10$), and of "supernova" extreme shock. All of the differential flux values are normalized with those at 10 GeV.

gases are assumed to be 36 and 42 eV, respectively. In addition, the ratio of the absorbed doses consumed for ionization of each gas is assumed to be proportional to its mass density, i.e. 75% and 25% of absorbed doses are used for ionization of hydrogen and helium gases, respectively, for the compound gas of cosmic abundance. Thus, the mean $W_0$ values for the compound gas is 37.5 eV. The attenuation lengths are estimated by least-square fittings for the depth range from 500 to 1500 g cm$^{-2}$. Contributions of protons as well as the sum of electrons and positrons are also drawn. In the same manner as the UN81 model, the attenuation lengths of the electromagnetic contributions are much longer than those of protons, and they are nearly equal to the total attenuation lengths. However, the attenuation lengths are longer than the corresponding data obtained from the UN81 model by approximately 20% for both protons and electromagnetic contributions.

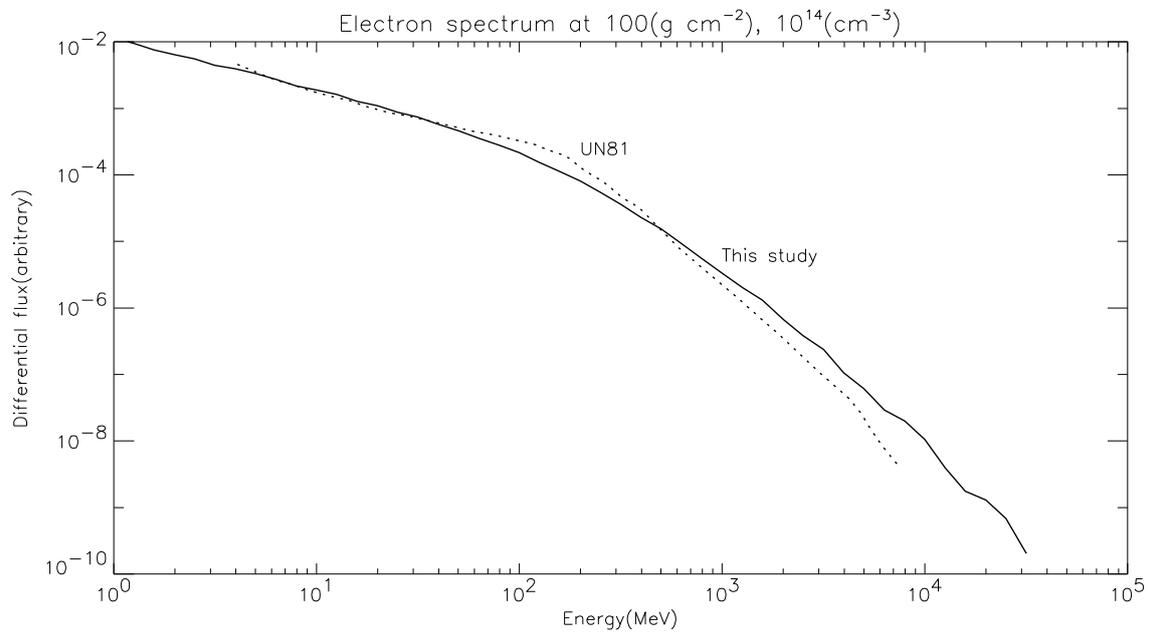

**Figure 3:** Electron spectrum obtained from PHITS for a hydrogen gas of $10^{14}$ cm$^{-3}$ at the depth of 100 g cm$^{-2}$, compared with the result of UN81 (dotted curve, taken from Figure 3a in UN81).

This discrepancy is attributed to the difference of secondary particle spectra produced by high-energy proton-proton inelastic scattering. **Figure 3** shows the electron spectrum at the depth of 100 g cm$^{-2}$ as obtained from PHITS simulation for $10^{14}$ cm$^{-3}$, compared with that of the UN81 model. It is evident that the PHITS spectrum is harder than that of UN81, and this harder spectrum results in the longer attenuation length of the electromagnetic contribution. This difference is probably due to the assumptions introduced in the inelastic scattering model in UN81 such as the concept of average inelasticity $\kappa$, and mono-energetic production of secondary particles. Conversely, PHITS simulates the high-energy proton-proton inelastic scattering using the JAM model, which solves the kinematics of each reaction considering the production of resonances and strings based on the Monte Carlo simulations.

Another possible reason for the discrepancy is the fact that the density dependences of proton fluxes obtained between the UN81 model and PHITS are different from each other. Proton fluxes are nearly independent of the gas density in the UN81 model, while they consistently become smaller with decrease of the density particularly at the deeper locations in the PHITS simulation. In the small domain of $10^{14}$ cm$^{-3}$, neutrons can travel through a certain mass thicknesses before they decay to deposit their energy at deeper locations when they decay into protons. In the large enough domain of $10^{10}$ cm$^{-3}$, on the other hand, the mean range of neutrons is very short in terms of the mass thickness, and hadrons are always suffered from the ionizations. Thus, the attenuation length of proton's ionization is supposed to be longer with the increase of the gas density, as suggested by the PHITS simulation.

### 3.2 Variety of attenuation lengths in several different settings

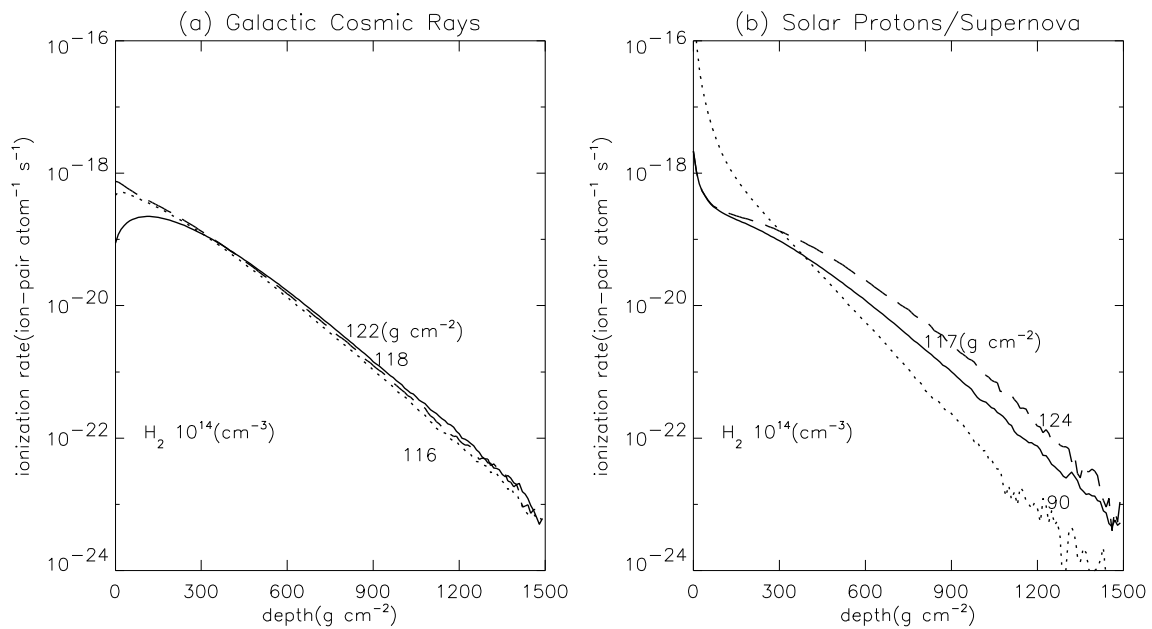

**Figure 4:** Ionization rates of a hydrogen gas of $10^{14}$ cm$^{-3}$. (a) Results from galactic cosmic rays of maximum ($W = 200$) and minimum ($W = 0$) modulations are shown by solid and dotted curve, respectively, with the result using the same condition of UN81 by a dashed curve. (b) Results from solar protons of strong shock ($m = 10$) and of weak shock ($m = 2$) are shown by solid and dotted curve, respectively, with the result from "supernova" extreme shock by a dashed curve.

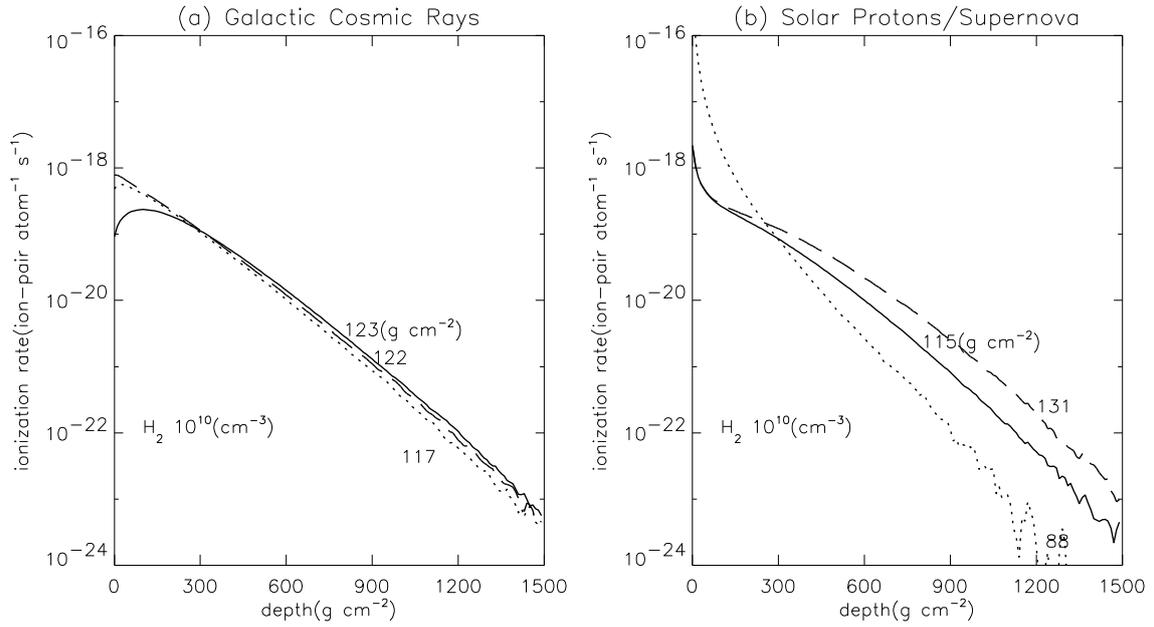

**Figure 5:** Same as Figure 4, except for the gas density of $10^{10}$ cm$^{-3}$.

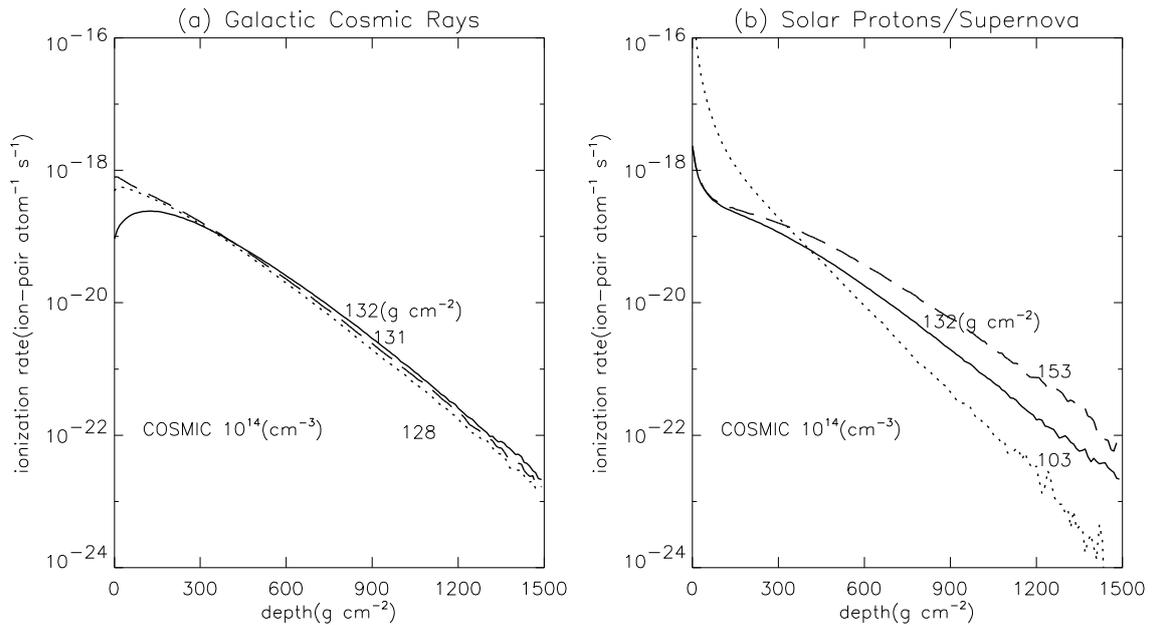

**Figure 6:** Same as Figure 4, except for the gas composition of cosmic abundance.

The obtained results of PHITS simulations for different values of the solar modulation index $W$ and shock Mach number $m$ are shown in **Figures 4 and 5**, with the attenuation lengths as estimated by least-square fittings for the depth range from 500 to 1500 g cm$^{-2}$. It is found that ionization profiles of solar minimum ($W = 0$) GCR is essentially the same as those of UN81 model, as expected from the similarity of the incident spectra shown in **Figure 1**. It is also found that strong shock ($m = 10$) solar protons and extreme shock "supernova" protons cause comparable and approximately 10% larger attenuation lengths with those of solar maximum ($W = 200$) GCRs, respectively. The weak shock ($m = 2$) solar protons cause only an order of magnitude smaller deep ionizations compared with other settings, with approximately 20% smaller attenuation length, and the significant ionization occurs mainly at 0-100 g cm$^{-2}$. This is naturally understood due to shorter ranges of primary particles as well as smaller contribution of secondary particles for the dominant lower energy incident particles.

The possible difference of the attenuation length due to different gas compositions are also briefly discussed. As shown in **Figure 6**, it is found that the attenuation lengths are approximately 10% larger in the gas of cosmic abundance than in pure hydrogen gas. This increase can be mainly explained by the smaller mass stopping power of the helium gas in comparison to the hydrogen gas. Combining with the discussions in Section 3.1 about the larger attenuation length of this study than of UN81, the much larger attenuation length further affects the minimum estimate of the size of dead zones in protoplanetary disks in previous studies. For example, the ionization rate can be approximately an order of magnitude larger at the depth of 1000 g cm$^{-2}$ than originally expected using the UN81 model.

Finally, the attenuation lengths due to the young Sun's solar protons from weak and strong shocks are ready to use for complicated simulations of protoplanetary disk's evolutions, by combining with the same formulation as shown by equation (A9) of Umebayashi and Nakano (2009). Note that stellar protons do not necessarily strike the surface of a circumstellar disk perpendicularly, and the projection effect can reduce the flux per unit disk area by a factor of 10 (Turner and Drake, 2009). In the present heliosphere, solar protons with hard spectra continue for only several hours with a significant beaming (Kataoka et al., 2014b). In the young Sun, however, the first good approximation would be a series of eruptive solar proton events with a frequent enough occurrence to be a steady state, with a random enough beaming directions to be uniform, so that the radial dependence of the young Sun's solar proton flux can be assumed to be proportional to the square root of the distance.

**Appendix: Solar protons from the young Sun**

Alfven radius of protosolar system is estimated by the balance between the kinetic energy of accretions and magnetic energy as

$$r_{\mathrm{A}} = \left( \frac{\mu^4}{2GM\dot{M}} \right)^{1/7} = 7.3 \left( \frac{\mu}{10^{36} \text{ G cm}^3} \right)^{4/7} \left( \frac{\dot{M}}{10^{-8} \text{ M}_\odot \text{ yr}^{-1}} \right)^{-2/7} \text{R}_\odot . \quad (A1)$$

where $\mu$ is the magnetic dipole moment of the central star, and $\dot{M}$ is the accretion rate. Here we assumed the largest possible young Sun's surface field of an order of kG. The magnetic field and Alfven speed at Alfven radius are estimated as

$$B_{\mathrm{A}} = \frac{\mu}{r_A{}^3} = 7.5 \left( \frac{\mu}{10^{36}\ \mathrm{G\ cm^3}} \right)^{-5/7} \left( \frac{\dot{M}}{10^{-8}\ \mathrm{M_\odot\ yr^{-1}}} \right)^{6/7}\ \mathrm{G}\ , \qquad (A2)$$

$$V_{\mathrm{A}} = \sqrt{\frac{B_A}{\rho_A}} = 3.0 \times 10^7 \left( \frac{\mu}{10^{36}\ \mathrm{G\ cm^3}} \right)^{1/14} \left( \frac{\dot{M}}{10^{-8}\ \mathrm{M_\odot\ yr^{-1}}} \right)^{-5/14}\ \mathrm{cm\ s^{-1}}\ , \qquad (A3)$$

where $\rho_{\mathrm{A}}$ is the mass density at the Alfven radius as determined from free fall accretion.

Differential number flux of shock-accelerated particles $dN/dE$ follows a power law (Blandford and Ostriker, 1978),

$$\frac{dN}{dE} \propto E^{-\gamma}\ , \qquad (A4)$$

and the power law index $\gamma$ is a function of Mach number $m$,

$$\gamma = 2 \frac{1 + m^{-2}}{1 - m^{-2}}\ . \qquad (A5)$$

The maximum energy of shock accelerated protons $E_{\max}$ can be estimated as a function of shock speed $V$, magnetic field $B$, and shock radius $R$, with a Bohm factor $\xi$ as follows,

$$E_{\max} = \frac{3}{20} \frac{1}{\xi} qVBR = 19 \left( \frac{\xi}{10} \right)^{-1} \left( \frac{V}{3 \times 10^7\ \mathrm{km/s}} \right) \left( \frac{B}{7.5\ \mathrm{G}} \right) \left( \frac{R}{7.3\ \mathrm{R_\odot}} \right)\ \mathrm{GeV}\ . \qquad (A6)$$

Combining the equations (A1), (A2), (A3), and (A6), the maximum energy $E_{\max}$ used in this paper can be described as a function of shock Mach number $m$ as follows.

$$E_{\max} = 1.9 \times 10^2 \left( \frac{\xi}{10} \right)^{-1} \left( \frac{m}{10} \right) \left( \frac{\mu}{10^{36}\ \mathrm{G\ cm^3}} \right)^{-3/14} \left( \frac{\dot{M}}{10^{-8}\ \mathrm{M_\odot\ yr^{-1}}} \right)^{13/14}\ \mathrm{GeV}\ . \qquad (A7)$$

Note that the $E_{\max}$ does not largely depend on the assumed magnetic dipole moment $\mu$, but nearly proportional to the mass loss rate $\dot{M}$.

**Acknowledgments:**
RK thanks Toshikazu Ebisuzaki and Shota Notsu for useful discussions on the possible space environment of T Tauri stars. This work was supported by JSPS KAKENHI Grant Number 26106006 and 15K13581.